\documentclass[paper,nofootinbib] {revtex4}
\usepackage[utf8x]{inputenc}
\usepackage{hyperref}
\usepackage{graphicx}
\usepackage{epsfig}
\usepackage{fancyhdr}
\usepackage{axodraw}
\usepackage{amsmath}
\usepackage{amsfonts}
\usepackage[font=small]{caption}
\usepackage{multirow}
\renewcommand{\phi}{\varphi}

\newcommand{\rhoomega}{$\rho$-$\omega$}
\newcommand{\ang}{\text{ang}}

\def\be{\begin{equation}}
\def\ee{\end{equation}}
\def\bea{\begin{eqnarray}}
\def\eea{\end{eqnarray}}

\begin{document}

\title{On the Spin of the $X(3872)$
}

%

\author{R. Faccini$^*$, F. Piccinini$^{\dag}$, 
A. Pilloni$^*$ and A.D. Polosa$^*$ 
} 
\affiliation{
$^*$Dipartimento di Fisica and INFN, `Sapienza' Universit\`a di Roma, 
\\
Piazzale A. Moro 2, Roma, I-00185, Italy
\;\;\;\; \\
$^\dag$INFN Pavia, Via A. Bassi 6, Pavia, I-27100, Italy
}

\begin{abstract}
Whether the much studied $X(3872)$ is an axial or tensor resonance makes 
an important difference to its interpretation.  A recent paper by the BaBar collaboration~\cite{delAmoSanchez:2010jr} raised the
viable hypothesis that it might be a $J^{PC}=2^{-+}$ state based on the $\pi^+ \pi^- \pi^0$ spectrum in the $X\to J/\psi\;\omega$ decays. 
Furthermore, the Belle collaboration published  the $\pi^+ \pi^- $  invariant mass and spin-sensitive angular distributions in  $X\to J/\psi\;\rho$ decays~\cite{Choi:2011fc}.

Starting from a general parametrization of the decay amplitudes 
for the axial and tensor quantum numbers of the $X$, we re-analyze
the whole set of available data.
The level of agreement of the two spin hypotheses with data is interpreted with a rigorous statistical approach based on Monte Carlo simulations in order to be able to combine all the distributions regardless of their different levels of sensitivity to the spin of the $X$. 

Our analysis returns a probability of  5.5\%  and  0.1\% for the agreement with data of the $1^{++}$ and $2^{-+}$ hypotheses, respectively, once we combine the whole information (angular and mass distributions) from both channels. 
On the other hand, the separate analysis of $J/\psi\; \rho$ (angular and mass distributions) and $J/\psi \;\omega$ (mass distribution) 
indicates that the $2^{-+}$ assignment is excluded at the 99.9\% C.L. by the former case, while the latter excludes at the same level the $1^{++}$ hypothesis. There are therefore indications that the two decay modes behave in a different way. \\ \newline

\end{abstract}


\maketitle

\thispagestyle{fancy}
\section{Introduction}
\label{introduction}
Although the $X(3872)$ resonance
is the most studied among the exotic $XYZ$ states, since its discovery in 2003, 
its quantum numbers have not   been definitively identified yet.
 
The CDF collaboration concluded from the analysis of the angular 
distributions and correlations of the $X$ decay products
that the possible quantum numbers are $J^{PC}= 1^{++}$ 
or $2^{-+}$~\cite{Abulencia:2006ma}. Similar results have been found very recently 
by the BELLE collaboration~\cite{Choi:2011fc}. 
In the latter paper
the $\pi^+ \pi^-$ invariant mass distribution is analyzed under the  
$1^{++}$ and $2^{-+}$ hypotheses, finding preference for 
the former  whereas no preferred assignment emerges if 
an interfering contribution with the isospin-violating decay 
$\omega \to \pi^+ \pi^-$ is added to the amplitude. 

The picture becomes more puzzling as one considers the analysis 
by the BaBar Collaboration of the decay 
$X \to J/\psi \;\pi^+ \pi^- \pi^0$~\cite{delAmoSanchez:2010jr}. 
The expected $3 \pi$ invariant mass distribution  agrees with data slightly better 
if the $2^{-+}$ signature is assumed. This result on the $3\pi$ spectrum was later confirmed in~\cite{Brazzi:2011fq}. 

In a recent paper~\cite{Hanhart:2011tn} the 
pion invariant masses in the decays $X \to J/\psi\; 2\pi$ 
and $X \to J/\psi \;3\pi$ are simultaneously analyzed with a combined fit, 
concluding that present data favor the $1^{++}$ assignment.  
In our view, the $2\pi$ invariant mass distribution is not  able to resolve 
the two hypotheses despite of the high statistics, and the way the fit was performed leads to the dilution of the sensitivity of the $3\pi$ channel. 

To improve the analysis of  all available data sensitive to the spin of the $X$  $i)$~we write the decay matrix elements for both the  $1^{++}$ and $2^{-+}$ hypotheses as given by enforcing Lorentz  invariance and parity considerations,
$ii)$~we give a functional dependency  on the decay momenta to the couplings introducing a length scale $R$ -- which is to be related to the finite size of the hadrons participating to the interactions,
$iii)$~we use the matrix elements of $\rho$ and $\omega$ decays to take into account the appropriate decay waves; we do not pursue the Blatt-Weisskopf  description as we find that all spin structure can appropriately be taken into account with no further approximations,
$iv)$~we perform a global fit to exploit the information contained in all the distributions available and we adopt a statistical approach appropriate when distributions with different sensitivities to the parameters of interest are combined.

\section{Matrix elements}
\label{sec:matrices}
The matrix elements describing the amplitudes $X\to J/\psi \,V$ (where $V=\rho,\omega$)
are obtained by Lorentz, gauge invariance and  parity considerations leading to the formulae reported below~\cite{Brazzi:2011fq}.

In the $X(1^{++})$ case we have
\begin{equation}
\label{eq:tensor1}
\langle \psi(\epsilon,p) V(\eta,q)|X(\lambda,P)\rangle = g_{1\psi V} \;\epsilon^{\mu\nu\rho\sigma}\;\lambda_\mu(P)\;\epsilon^*_\nu(p)\;\eta^*_\rho(q)\;P_\sigma
\end{equation}
the polarization vectors carry a complex conjugation when referred to final states.

In the   $2^{-+}$ case we have a more complicated structure
\begin{equation}
\label{eq:tensor2}
\langle \psi(\epsilon,p) V(\eta,q)|X(\pi,P)\rangle = g_{2\psi V} \;T_A + g_{2\psi V}^\prime \;T_B
\end{equation}
where $\pi$ is the polarization tensor for a spin two particle with mass\footnote{
The sum over polarizations is 
\begin{equation}
\label{propg}
\sum_{\rm pol}\pi_{\mu\nu}(k)\pi^*_{\alpha\beta}(k) = \frac12 T_{\mu\alpha} T_{\nu\beta} + \frac12 T_{\mu\beta} T_{\nu\alpha} - \frac13 T_{\mu\nu} T_{\alpha\beta}
\end{equation}
with $T_{\mu\nu} = -g_{\mu\nu} + k_\mu k_\nu / m^2$ and $k^2=m^2$.
}
and a standard notation is used for the remaining polarization vectors. We find that  $T_{A}$ and $T_{B}$ are given by
\be
T_A=\epsilon^{*\alpha}(p)\:\pi_{\alpha\mu}(P)\;\epsilon^{\mu\nu\rho\sigma}\;p_\nu\;q_\rho\;\eta^*_\sigma(q)- (\epsilon,p \leftrightarrow   \eta, q )
\label{four}
\ee
and 
\be
T_B=Q^\alpha\:\pi_{\alpha\mu}(P)\;\epsilon^{\mu\nu\rho\sigma}\;P_\nu\;\epsilon^*_\rho(p)\;\eta^*_\sigma(q)
\ee
where $Q=p-q$ and $P=p+q$.  

The coupling  $g_{1\psi V}$ is real whereas the couplings  $g_{2\psi V}$  and  $g_{2\psi V}^\prime$  are separately real but can have a complex relative phase. This is due to the fact that, on the basis of Lorentz invariance and parity conservation, we can indeed write three terms $T_{A}^{(\lambda)},T_{B}^{(\lambda)}$ and $T_{C}^{(\lambda)}$  where $\lambda$ labels  one out of the $3\times3\times5$ polarization combinations which define $T_{A},T_{B},T_{C}$.  $T_{C}$ is the same as $T_{A}$ but with a plus relative sign between the two terms on the rhs of~\eqref{four}. Only two out of these three terms are linearly independent (say $T_{A}$ and $T_{B}$) for
\be
\left|\sum_{\lambda=1}^{3\times 3\times 5} T_{A}^{(\lambda)}T_{C}^{(\lambda)*}\right|^{2}=\sum_{\lambda} |T_{A}^{(\lambda)}|^{2} \sum_{\lambda} |T_{C}^{(\lambda)}|^{2}
\ee
which is the equality of the Schwartz inequality: this  holds  if $z_{1} T_{A}= z_{2}T_{C}$ where $z_{1}$ and $z_{2}$ are two complex numbers both different from zero.  Thus we can exclude $T_{C}$ and retain  $T_{B}$ and, in general,  $z_{1}T_{A}$ to characterize the decay amplitude.

In Refs.~\cite{Choi:2011fc,delAmoSanchez:2010jr,Hanhart:2011tn} the $P$-wave fit functions contain a Blatt-Weisskopf angular momentum barrier factor of the form $\left(1+R^2 q^{*2}\right)^{-1/2}$, where $q^*$ is the $X$ decay 3-momentum. The value of $R$ cannot be
extracted from data since the $P$-wave distribution will approach the $S$-wave distribution in the limit $R\to\infty$, so that if we let $R$ free, the fit will not converge.
On the other hand, in our discussion we do not need any barrier 
factor the decay wave being dictated by the expressions of the matrix elements. We instead take into account the finite size of the $X$ (and of $V$ and $J/\psi$ as well)
introducing  a `polar' form factor, namely
\begin{equation}
\label{polarfactor}
g \to \frac{ g }{(1+R^2 q^{*2})^{n}}
\end{equation}
where $g$ stands in general for $g_{1\psi V}$, $g_{2\psi V}$ and $g_{2\psi V}^\prime$. We tested the values $n=1$ and $n=2$ (the latter coinciding with the  Fourier transform of a an exponential $g(r)\sim\exp(-r/R)$ strong charge distribution). Both the fitting functions turn out to be rather effective at improving our results, with no significative change for the
the two choices of $n$.
We also  underscore that $g_{1\psi V}$ (regulating the $S$-wave decay) is assumed to have the same polar behavior since Eq.~\eqref{polarfactor} does not concern any orbital angular momentum considerations.
The size parameters $R_{J}$ will eventually be fitted from data.

As for the $\rho$ and $\omega$ decay amplitudes, we use
\begin{equation}
\label{eq:tensorrho}
 \langle \pi^+(p) \pi^-(q) |\rho\left(\epsilon,P\right)\rangle = g_{\rho 2\pi} \,\epsilon \cdot p
\end{equation} 
which describes a $P$-wave decay (the square modulus of this matrix element is  $g^2_{\rho 2\pi}$ times the decay momentum squared). For the $\omega$ we have
\begin{equation}
\label{eq:tensoromega}
 \langle \pi^+(p) \pi^-(q) \pi^0(r) |  \omega\left(\epsilon,P\right) \rangle =  g_{\omega 3\pi}\, \epsilon^{\mu\nu\rho\sigma} \epsilon_\mu p_\nu q_\rho r_\sigma 
\end{equation}

The last two couplings will  simply be written in terms of the partial widths $\Gamma(\rho\to 2\pi)$ and
$\Gamma(\omega\to 3\pi)$, as shown in the next section.

\section{Decay Widths}
\label{sec:widths}
We have to calculate 
the partial widths $\Gamma(X\to J/\psi\;\pi^+\pi^-)$ and 
$\Gamma(X\to J/\psi\;\pi^+\pi^-\pi^0)$. In what follows we will neglect the {\rhoomega} mixing since we demonstrate in Appendix~\ref{sec:appendix} that it does not alter significantly the results.

The partial widths in the narrow width approximation are~\cite{Brazzi:2011fq}
\begin{equation}
\label{eq:gamma2pigrezza}
\begin{split}
\Gamma(X\to J/\psi\;\pi^+\pi^-) &=\frac{1}{2J+1}\frac{1}{48\pi m^2_X}\int\;ds\;\sum_{\substack{{\rm pol}}}|\langle \psi\;\rho(s)|X\rangle|^2p^*(m^2_X,m^2_\psi,s)\\
&\quad \times \frac{1}{\pi}\frac{1}{(s-m_\rho^2)^2+(m_\rho\Gamma_\rho)^2} \int d\Phi^{(2)} \sum_{\substack{{\rm pol}}} |\langle \pi^+ \pi^-|\rho(s) \rangle|^2
\end{split}
\end{equation}
where by $d\Phi^{(2)}$ we mean the 2-body phase space measure. 
The decay momentum $q^{*}$ in the matrix element $\langle \psi\;\rho(s)|X\rangle$ 
coincides with $q^{*}\equiv p^*(m^2_X,m^2_\psi,s)$; similarly for the width in $J/\psi\; \omega$ discussed below.

The sum over polarizations in \eqref{eq:tensorrho}, simply yields
\begin{equation}
\label{eq:polarizrho}
\sum_{\substack{{\rm pol}}} |\langle \pi^+ \pi^- |\rho(s) \rangle|^2 = g_{\rho 2\pi}^2\; p^*(s,m^2_\pi,m^2_{\pi})^2
\end{equation}
Finally, we can eliminate the coupling by evaluating the \eqref{eq:polarizrho} on the mass-shell and by relating it to the partial width $\Gamma \left(\rho\to\pi\pi\right)$
\begin{equation}
\label{eq:couplingrho}
g^2_{\rho 2\pi}= 6 m^2_\rho\;\Gamma(\rho\to\pi\pi)\;\frac{4\pi}{p^*(m_\rho^2,m^2_\pi,m^2_\pi)^3}
\end{equation}

Inserting the above expressions \eqref{eq:polarizrho}, \eqref{eq:couplingrho}
into \eqref{eq:gamma2pigrezza} gives
\begin{equation}
\label{eq:gamma2pi}
\begin{split}
\Gamma(X\to J/\psi\;\pi^+\pi^-) &=\frac{1}{2J+1}\frac{1}{8\pi m^2_X}\int\;ds\;\sum_{\substack{{\rm pol}}}|\langle \psi\;\rho(s)|X\rangle|^2p^*(m^2_X,m^2_\psi,s)\\
&\quad\times\frac{1}{\pi}\frac{m_\rho\Gamma_\rho\; \mathcal{BR}(\rho\to\pi\pi)}{(s-m_\rho^2)^2+(m_\rho\Gamma_\rho)^2}\frac{m_\rho}{\sqrt{s}}\left(\frac{p^*(s,m^2_\pi,m^2_{\pi})}{p^*(m_\rho^2,m^2_\pi,m^2_\pi)}\right)^3\\
\end{split}
\end{equation}

Similarly, for the $\omega$ we obtain
\begin{equation}
\label{eq:gamma3pigrezza}
\begin{split}
\Gamma(X\to J/\psi\;\pi^+\pi^-\pi^0) &=\frac{1}{2J+1}\frac{1}{48\pi m^2_X}\int\;ds\;\sum_{\substack{{\rm pol}}}|\langle \psi\;\omega(s)|X\rangle|^2p^*(m^2_X,m^2_\psi,s)\\
&\quad \times \frac{1}{\pi}\frac{1}{(s-m_\omega^2)^2+(m_\omega\Gamma_\omega)^2} \int d\Phi^{(3)} \sum_{\substack{{\rm pol}}} |\langle \pi^+ \pi^- \pi^0|\omega(s) \rangle|^2
\end{split}
\end{equation}

Summing over the $\omega$ polarizations 
\begin{equation}
\sum_{\substack{{\rm pol}}} |\langle \pi^+ \pi^- \pi^0|\omega\left(s\right)\rangle|^2 = \frac{g^2_{\omega 3\pi}}{s^3}\frac{s}{4} \left[\left( m^2_0 + s - 2 \sqrt{s} \;\omega \right)\left(\omega^2-m^2_0-4x^2\right)
 - 4 m^2_+ \left(\omega^2 - m^2_0\right)\right] \equiv g^2_{\omega 3\pi} \mathcal{M}\!\left(\sqrt{s}\right)
\label{eq:polarizomega}
\end{equation}
where $\omega = E_{\pi^0}$, $x = \frac12 \left(E_{\pi^+}-E_{\pi^-}\right)$ and  $m_0 = m_{\pi^0}$,  $m_+ = m_{\pi^+} = m_{\pi^-}$.
An adimensional coupling has been  formed by  substituting $g^2_{\omega 3\pi} \to \frac{1}{s^3} g^2_{\omega 3\pi}$~\footnote{This rescaling is arbitrary and in part relies on the narrowness of the $\omega$.
Avoiding the introduction of the $1/s^3$ term, the separate fit of $3\pi$ does not change significantly (for example, with $n=1$, the $2^{-+}$ is unchanged, the $1^{++}$ gets worse from $\chi^2=9.9 \to 11.1$).
In the combined fit of invariant mass distributions both hypotheses become a bit worse  (again for $n=1$, $2^{-+}$: $\chi^2 = 17.7\to 18.4$;      
 $1^{++}$: $\chi^2 = 25.2\to 26.2$) so that the $\Delta\chi^2$ remains almost unchanged. See Sec.~\ref{sec:statistics}.}.
We eliminate the coupling by evaluating \eqref{eq:polarizomega} on the mass-shell
\begin{equation}
\label{eq:couplingomega}
g^2_{\omega 3\pi}= 6 m_\omega\;\Gamma(\omega\to 3\pi)\left( \int d\Phi^{(3)} \mathcal{M}\!\left(m_\omega\right) \right)^{-1}
\end{equation}

Inserting \eqref{eq:polarizomega}, \eqref{eq:couplingomega} into  \eqref{eq:gamma3pigrezza} we obtain
\begin{equation}
\label{eq:gamma3pi}
\begin{split}
\Gamma(X\to J/\psi\;\pi^+\pi^-\pi^0) &=\frac{1}{2J+1}\frac{1}{8\pi m^2_X}\int\;ds\;\sum_{\substack{{\rm pol}}}|\langle \psi\;\omega(s)|X\rangle|^2p^*(m^2_X,m^2_\psi,s)\\
&\quad\times\frac{1}{\pi}\frac{m_\omega\Gamma_\omega\; \mathcal{BR}(\omega\to3\pi)}{(s-m_\omega^2)^2+(m_\omega\Gamma_\omega)^2} \frac{{\Phi^{(3)\prime}}(\sqrt{s},m_0,m_+,m_+)}{{\Phi^{(3)\prime}}(m_{\omega},m_0,m_+,m_+)}
\end{split}
\end{equation}
where 
\begin{equation}
{\Phi^{(3)\prime}}(\sqrt{s},m_0,m_+,m_+) =\frac{1}{32\pi^{3}} \int_{m_0}^{\omega_m} d\omega \int_{x_-}^{x_+} dx \frac{1}{4s^2} \left[\left( m^2_0 + s - 2 \sqrt{s} \;\omega \right)\left(\omega^2-m^2_0-4x^2\right) - 4 m^2_+ \left(\omega^2 - m^2_0\right)\right] 
\end{equation}
with $\omega_m=(m^2_0-4m^2_++s)/(2 \sqrt{s})$ and
\begin{equation}
x_\pm=\pm\frac12 \sqrt{\frac{\left(\omega^2-m_0^2\right) \left(\omega_m-\omega \right)\sqrt{s}}{4m^2_+ +\left(\omega_m-\omega\right)\sqrt{s}}}
\end{equation}
to be compared to the notations used in~\cite{Brazzi:2011fq}. 

In formulae \eqref{eq:gamma2pi} and  \eqref{eq:gamma3pi} angular correlations are not taken into account for we factorize matrix elements. On the other hand, the only way to consider both off-shellness and angular correlations is to compute the full matrix element for the $1\to 5$ decay
\begin{equation}
\label{eq:gamma5bodies}
\begin{split}
\Gamma\left(B\to K\,X \to K\,J/\psi\,\rho \to K\,l^+l^-\,\pi^+\pi^-\right) &=\frac{1}{2 m_B} \int\;\prod_{i=1}^5 \frac{d^3p_i}{(2\pi)^3 2E_i} (2\pi)^4 \delta\left(p_B - \sum_i p_i\right) \\
&\quad\times\sum_{\substack{{\rm pol}}} |\langle K\,l^+l^-\,\pi^+\pi^-|B \rangle|^2
\end{split}
\end{equation}

The matrix element can be decomposed as 
\begin{equation}
\begin{split}
 \langle K\,l^+l^-\,\pi^+\pi^-|B \rangle &= \langle l^+l^- | \psi \rangle\, \frac{1}{p^2_\psi - m^2_\psi + i m_\psi \Gamma_\psi}\, \langle \pi^+ \pi^-  |\rho \rangle \,\frac{1}{p^2_\rho - m^2_\rho + i m_\rho \Gamma_\rho} \\
 &\quad \times\langle \psi\,\rho |X \rangle\, \frac{1}{p^2_X - m^2_X + i m_X \Gamma_X}\, \langle K \,X |B \rangle
 \end{split}
 \label{eq:fiveb}
\end{equation}

We already gave a form to $\langle \psi \,\rho|X\rangle$ and $\langle \pi^+ \pi^-|\rho\rangle$ in Sec.~\ref{sec:matrices} for both $1^{++}$ and $2^{-+}$. Moreover
\begin{align}
 \langle l^+ \, l^- |\psi \rangle &= (\epsilon_\psi)_\beta \;\bar{u}(l^-) \gamma^\beta v(l^+)\\
\langle \pi^+ \, \pi^- |\rho \rangle &= \epsilon_\rho \cdot p_{\pi^+}\\
\langle X \,K |B \rangle &= \left\{\begin{aligned}&\epsilon^*_X \cdot p_K & \mbox{for }&1^{++}\\
&(\pi^*_X)_{\delta\phi} (p_K)^\delta (p_K)^\phi & \mbox{for }&2^{-+}\end{aligned}\right.
\end{align}

The sum over polarizations of inner legs returns the usual numerators of spin-1 and spin-2 propagators. The expressions obtained are not reported here because of their algebraic complexity~\footnote{These are available upon request in form of Fortran routines.}. All matrix elements but the $\langle \psi\,\rho|X(2^{-+})\rangle$ can be written in terms of one coupling only times a scalar  function of momenta and polarizations. 

The widths are expressed in terms of sums of products of couplings times scalar functions of momenta. Products of couplings are absorbed within the fit parameters $r_{J}^{\ang}$ see  Sec.~\ref{sec:fit}.

Since the integral in Eq.~\eqref{eq:gamma5bodies} has to be evaluated numerically, we use  Monte Carlo techniques. The integration on the 5-body phase space is carried out using importance sampling on the Breit-Wigner peaks.
As a further option of the calculation, unweighted decay configurations can be generated according to full matrix element weights.   

As usual, in the calculations we use the so-called \emph{comoving width}, 
i.e. we rescale the width in the denominators of Eqs.~\eqref{eq:gamma2pi}, \eqref{eq:gamma3pi} and \eqref{eq:fiveb}  according to the standard prescription $m\Gamma \to  (s/m)\Gamma$.

\section{Data re-analysis}
\label{sec:fit}

\begin{figure}[!h]	
\begin{centering}
\includegraphics[width=10truecm]{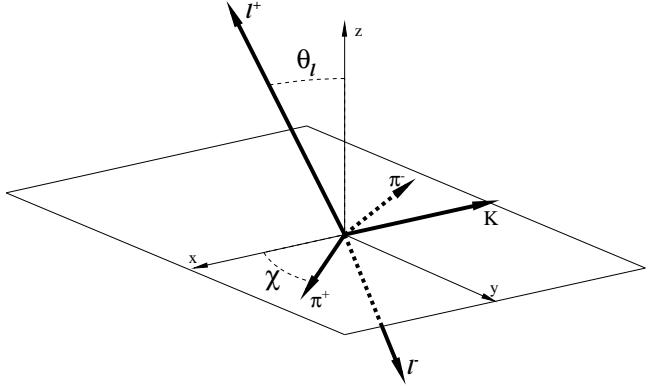}
\caption{Definition of the angles in  $X \to J/\psi\; \pi^+\pi^-$~\protect\cite{Choi:2011fc}.}
\label{fig:rosner_angles}
\end{centering}
\end{figure}

In order to extract information on the spin of the $X$ particle, we re-analyze the $B\to X K$ data published in  Ref.~\cite{Choi:2011fc} (for $X\to J/\psi\; \pi^+\pi^-$) and in Ref.~\cite{delAmoSanchez:2010jr} (for $X\to J/\psi\; \pi^+\pi^-\pi^0$).
In particular, in the $X \to J/\psi\; \pi^+\pi^-$ sample  we consider the di-pion invariant mass ($m_{2\pi}$) and the angles defined in the $X$ rest frame in Ref.~\cite{Rosner:2004ac} and described in Fig~\ref{fig:rosner_angles}:  the angle between the $J/\psi$ and  the direction opposite to $K$ ($\theta_X$), 
 the angle between the $\pi^+$ and the direction opposite to  $K$ ($\chi$), and 
 the angle between the $l^+$ produced by the decay of the $J/\psi$ and the $z$-axis of a coordinate system where the $x$-axis is the direction opposite to  the $K$ and the $y$-axis is the component of $\pi^+$ orthogonal to $K$  ($\theta_l$).
In the $X \to J/\psi\; \pi^+\pi^-\pi^0$, instead, only the three pions invariant mass is considered ($m_{3\pi}$). 

With such distributions at hand the parameters of the model  we have sensitivity upon are
\begin{itemize}
\item the radii $R_J$ ($J=1,2$) as defined in Eq.~\eqref{polarfactor}, to which only the invariant mass distributions are sensitive
\item the relative amplitudes and phases of the two contributions in case $J=2$ ($g_{2\psi V}$ and $g^\prime_{2\psi V}$ in Eq.~\eqref{eq:tensor2}). 
We redefine
\begin{equation}
\begin{aligned}
g_{2\psi \xi} &= r_{2\xi} \cos \frac{\theta_\xi}2, & g'_{2\psi\xi} &= r_{2\xi} \sin \frac{\theta_\xi}{2}e^{i\phi_\xi}\\
 \end{aligned}
\end{equation}
where $\xi$ can be $\rho,\; \omega$ or $\ang$, depending on whether we are fitting the $m_{2\pi}$, $m_{3\pi}$, or angular distributions respectively. The parameters $\theta_\rho$ and $\phi_\rho$ satisfy $\theta_\rho=\theta_{\ang}$ and $\phi_\rho=\phi_{\ang}$ since  in the $\rho$ channel we have information on both mass spectrum and angular distributions.  
Only the angular distributions are sensitive to the $\theta$ and $\phi$ parameters
\item the overall normalizations, $r_{2\xi}$ and $r_{1\xi}$ ($=g_{1\psi\rho}$ in Eq.~(\eqref{eq:tensor1})). It has to be noted that the normalization depends on the  distributions being studied
\end{itemize}
Given the different sensitivities of the mass and angle distributions and the computational complexity of the angular fits, the combined fits are performed in three steps: $i)$ the invariant mass distributions are fitted letting all parameters float, $ii)$ the fits to the three angular distributions are performed fixing the $R_J$ parameters to the results obtained in the previous fits, $iii)$ the invariant mass fits are repeated by fixing the $\theta_\rho$ and $\phi_\rho$ distributions to the results of the angular fits. $\phi_\omega$ and $\theta_\omega$ are always set to zero.

Mass and widths of the $\rho$, $\omega$, and $X$ are fixed.
To account for the $X$ width in the invariant mass distributions, we extract randomly the values of the mass of the $X$ from a Breit-Wigner centered at $m_X = 3872\,\textrm{MeV}$ with $\Gamma_X = 1.7\,\textrm{MeV}$~\cite{review}, accepting only those
values kinematically consistent with the decay of interest. 

Finally, since as described  in Appendix~\ref{sec:nequal2} the invariant mass fits return consistent results with either the  $n=1$ or the $n=2$ hypotheses (where $n$ is defined in Eq.~\eqref{polarfactor}) we will only report results with $n=1$. 

\begin{figure}[!h]
\advance\leftskip0cm
\includegraphics[width=8.4truecm]{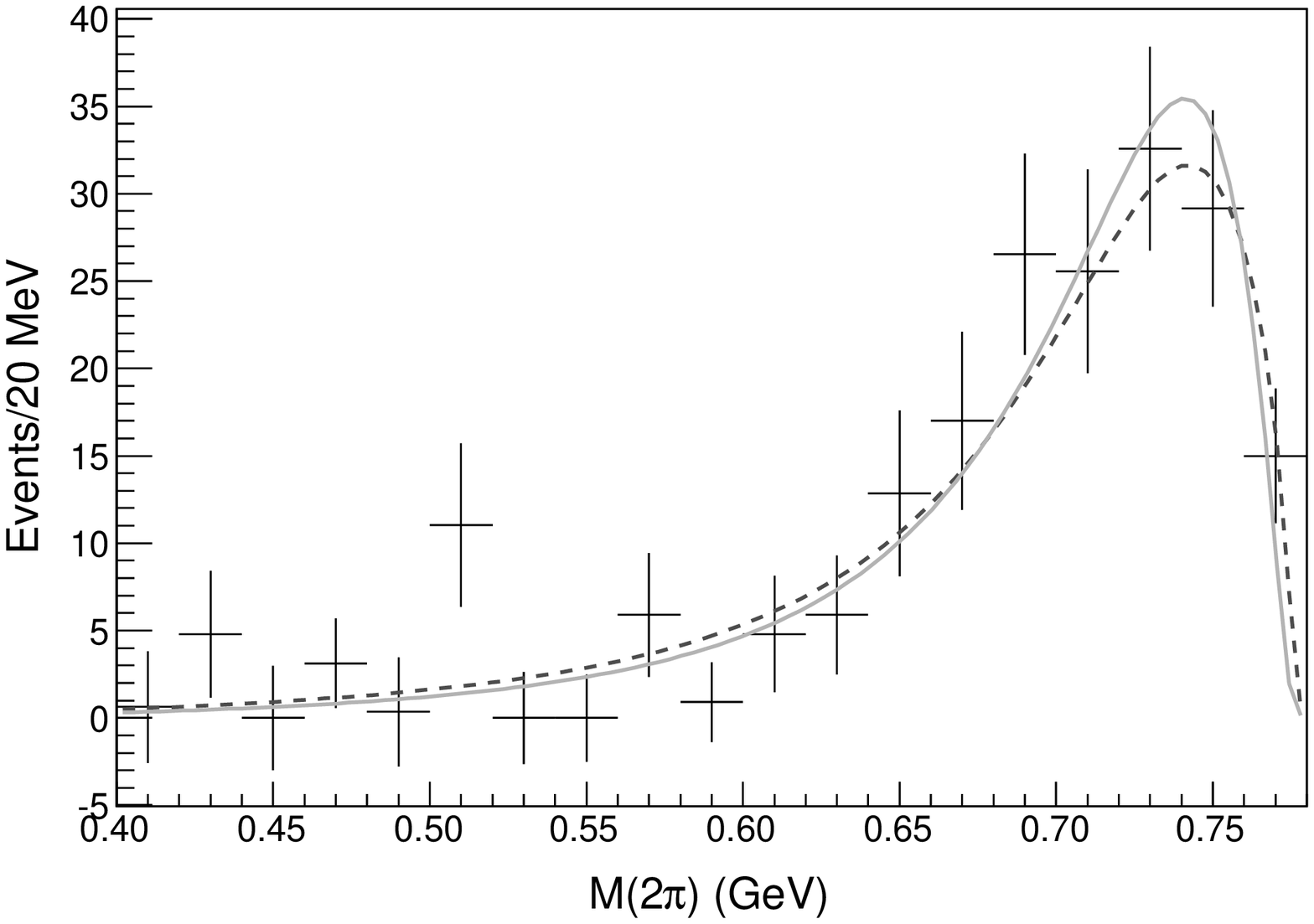}
\includegraphics[width=8.4truecm]{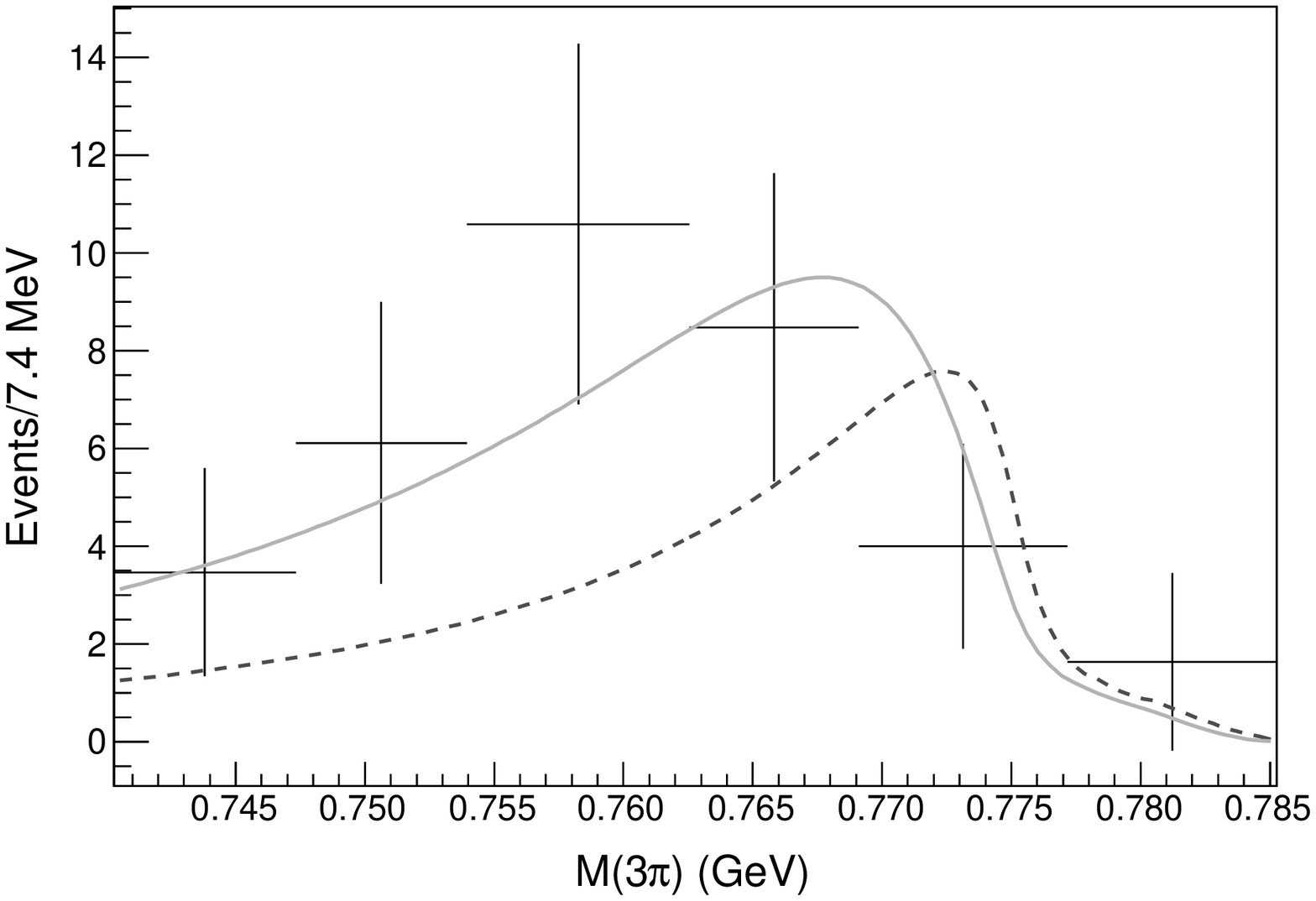}
\caption{Fit to the $m_{2\pi}$ (left) and $m_{3\pi}$ (right) distributions as described in the text, with the $n=1$ model. The dashed curve refers to the $1^{++}$ hypothesis whereas the solid one is for the $2^{-+}$.}
\label{fig:fitn1}
\end{figure}

\begin{figure}[!h]	
\includegraphics[width=8.0truecm]{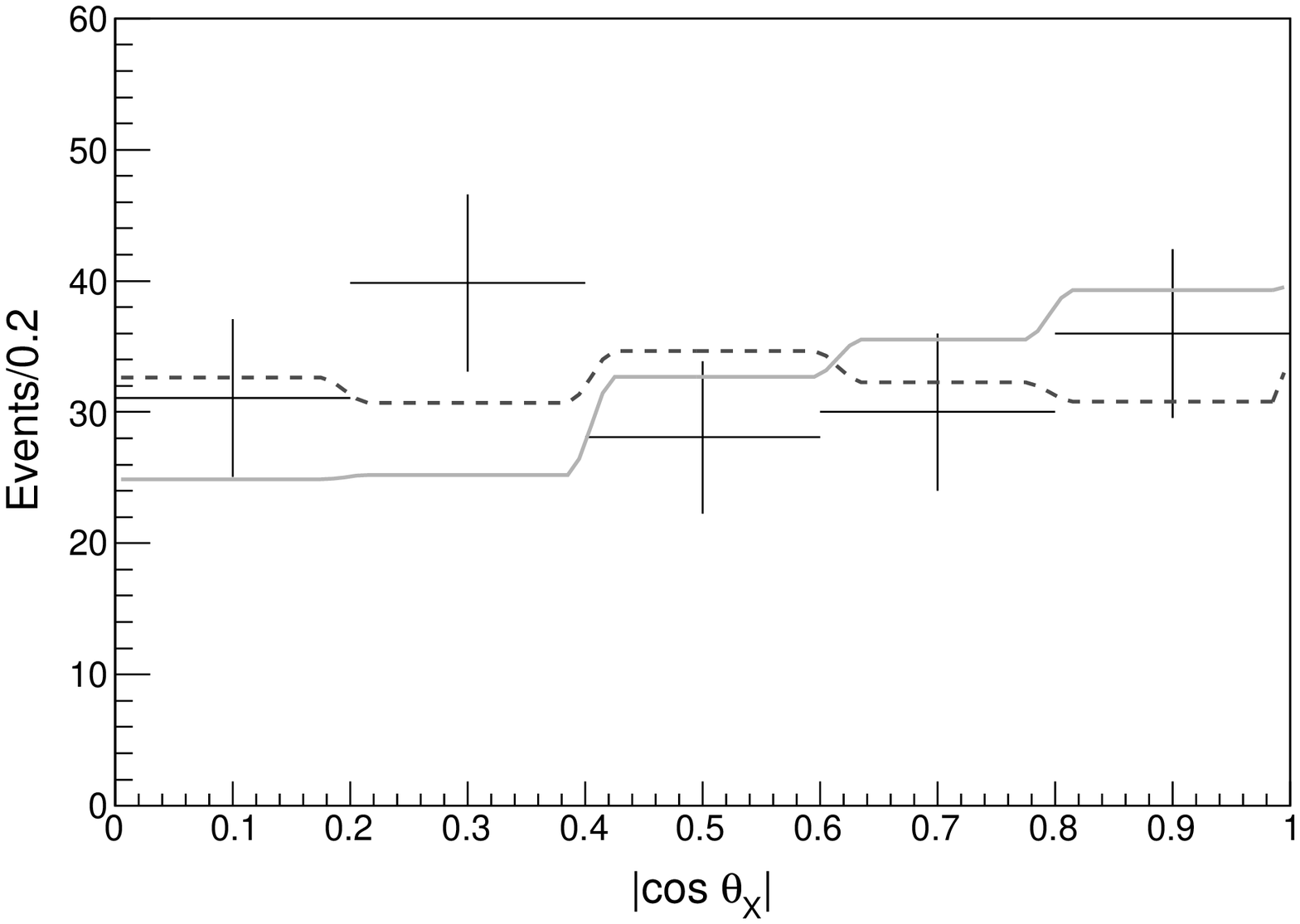}
\includegraphics[width=8.0truecm]{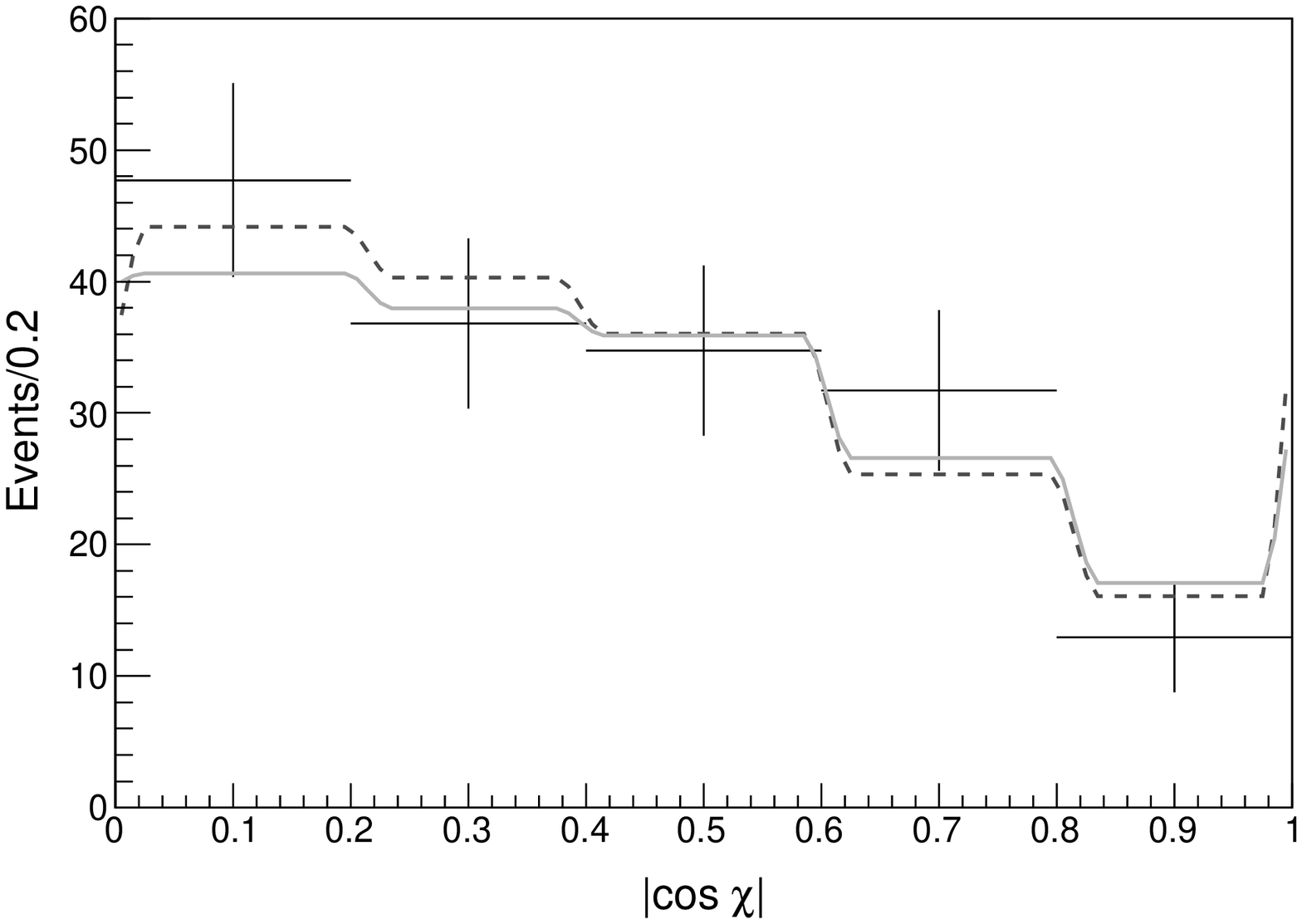}
\includegraphics[width=8.0truecm]{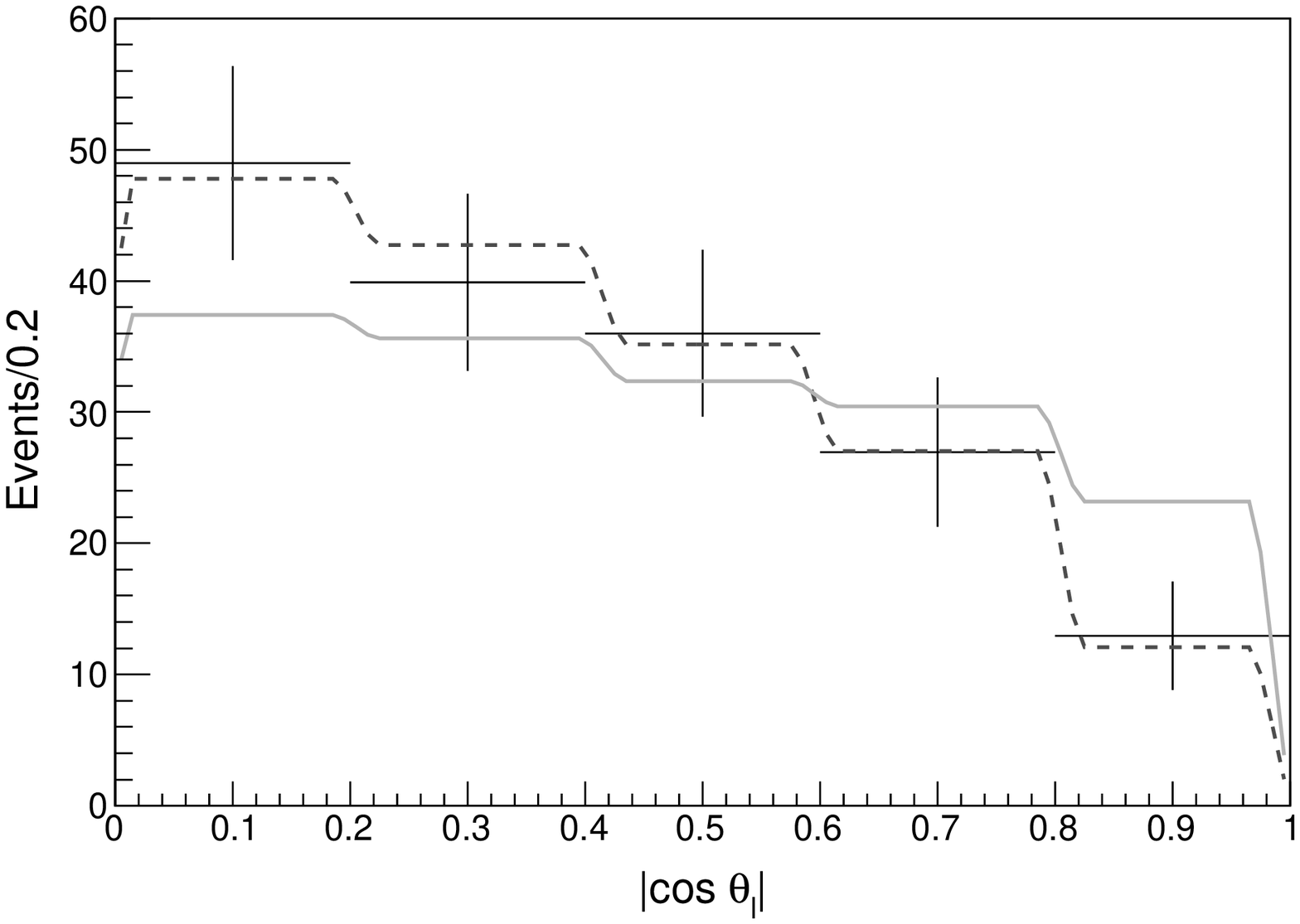}
\caption{Fit to the angular distributions in   $X \to J/\psi\, \pi^+\pi^-$ decays as described in the text. The dashed curve refers to the $1^{++}$ hypothesis whereas the solid one is for the $2^{-+}$.}
\label{fig:angles}
\end{figure}

The resulting fits are shown in Figs.~\ref{fig:fitn1} and ~\ref{fig:angles} and the fitted parameters are summarized in Tab.~\ref{tab:couplings}.
It is interesting to note that the radii, which are the  fit parameters with a physical content, have reasonably small errors
and get values consistent with $1$~fm, the size scale of a standard hadron. Here it is hard to judge if we are probing the size of the $X$, in which case we would say that the results obtained for $R$  disfavor a large ($\sim 10$~fm) loosely bound molecule~\cite{loose}, or the size of the interaction making the $X$ decay into $J/\psi+V$. For sure the loosely bound molecule  is requested to have a large wave function at the origin to make such a decay possible~\footnote{Moreover we remind that loosely bound molecules are very difficult to be produced with high cross sections at hadron colliders~\cite{noigrin} if not invoking final state interactions effect~\cite{artoise}.}. On the other hand in any compact multiquark model the decay into $J/\psi$ does not require special conditions.

\begin{table}[b]
\begin{tabular}{||c|c|c||} \hline
 & $1^{++}$ (dashed curve) & $2^{-+}$ (solid curve) \\ \hline
$r_{J\rho}$ & $0.089 \pm 0.006$ a.u. & $0.69 \pm 0.13$ a.u.\\ \hline
$r_{J\omega}$ & $0.0026\pm 0.0003$ a.u.  & $0.030\pm 0.016$ a.u. \\ \hline
$r_{J\ang}$ & $1.32 \pm 0.4$ a.u. & $1.03 \pm 0.04$ a.u.\\ \hline
$\theta_{\rho}$ & -  & $(254\pm 16)^\circ$ \\ \hline
$\phi_{\rho}$ & -  & $(14\pm 60)^\circ$ \\ \hline\hline
$R_J$ & $1.6 \pm 0.3\,\textrm{GeV}^{-1}$ & $5.6 \pm 0.8\,\textrm{GeV}^{-1}$\\\hline\hline
$\chi^2 / \textrm{DOF}$ & $31.8 / 36$  & $37.3 / 33$\\ \hline
$P(\chi^2) $ & $67\%$  & $28\%$\\ \hline
\end{tabular}
\caption{Fit results for the two $J^{PC}$ assignments.
}
\label{tab:couplings}
\end{table}

\subsection{Statistical interpretation}
\label{sec:statistics}

The na\"{i}ve approach to evaluate the likelihood of the two spin hypotheses would be to consider the $\chi^2$ of the fits, obtained by adding the contributions from all the distributions, and to compare them with the number of degrees of freedom. In our case this would return a $\chi^2/ \textrm{DOF}=31.8/36$ for the $1^{++}$ hypothesis and $\chi^2/ \textrm{DOF}=37.3/33$ for the $2^{-+}$ hypothesis. The probability of obtaining a worse value is $P(\chi^2)= \int_{\chi^2}^\infty {\rm PDF}(x,N_{\rm DOF}) dx=67\%$ and $28\%$ respectively, {\it i.e.}, both hypotheses seem to fit  data well.

Nonetheless this conclusion would not consider two effects.  First, the overall good agreements is caused by the fact that we are simultaneously considering   the distributions which favor the $1^{++}$ hypothesis ($\theta_X, \chi, \theta_l$) and the one which favors the $2^{-+}$ hypothesis ($m_{3\pi}$). This can be qualitatively seen in the fit results (Figs.~\ref{fig:fitn1} and~\ref{fig:angles}) and in the $P(\chi^2)$ values obtained on each fitted distribution as listed in Tab.~\ref{tab:toyangular}. Then, a large number of degrees of freedom are not sensitive to the spin of the $X$ and therefore wash out the overall $\chi^2$. This is the case for the $m_{2\pi}$ distribution, which is sensitive only to the radii, but not to the spin.

To quantify this latter statement and  develop a sounder statistical analysis, we performed a Monte Carlo (MC) study. We generate $N$ data samples with the same number of  events as the experimental samples and take into account  the
background, the statistical fluctuations, and the uncertainties in  the model 
parameters. 
The simulations can be performed either by making the hypothesis that the $X$ is truly a $1^{++}$  state or a $2^{-+}$ state.
 
As the starting point of the simulation for a given $J$ and $n$ hypothesis we take the corresponding model extracting sample by sample  the parameters according to the result of the combined fit to the data.
The  invariant mass MC samples are generated  by filling  the bins $b_i$ according to Poisson distributions with mean values $\mu_i$ corresponding, bin per bin, to the sum of the values expected from the model and the experimental background. The angular pseudodata are generated as MC unweighted events. 

After the  extraction this background is treated in the same way as the data, subtracted in the case of the invariant mass distributions and accounted for in the fits in the case of the angular distributions. 

The $N$ data samples obtained with this procedure are then analyzed with the same fitting function and statistical analysis as the experimental data.

\begin{figure}[!h]	
\includegraphics[width=8truecm]{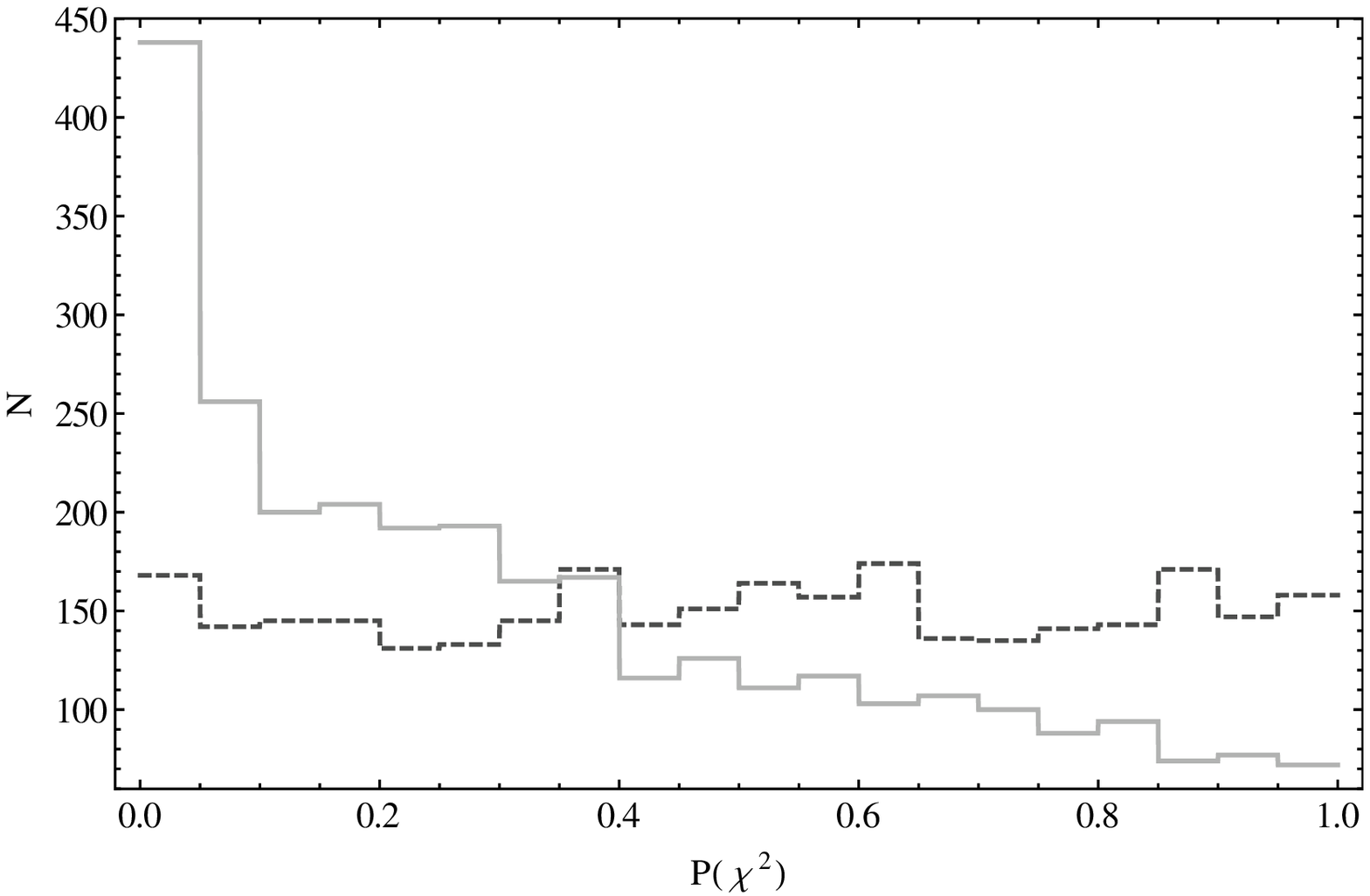}
\includegraphics[width=8truecm]{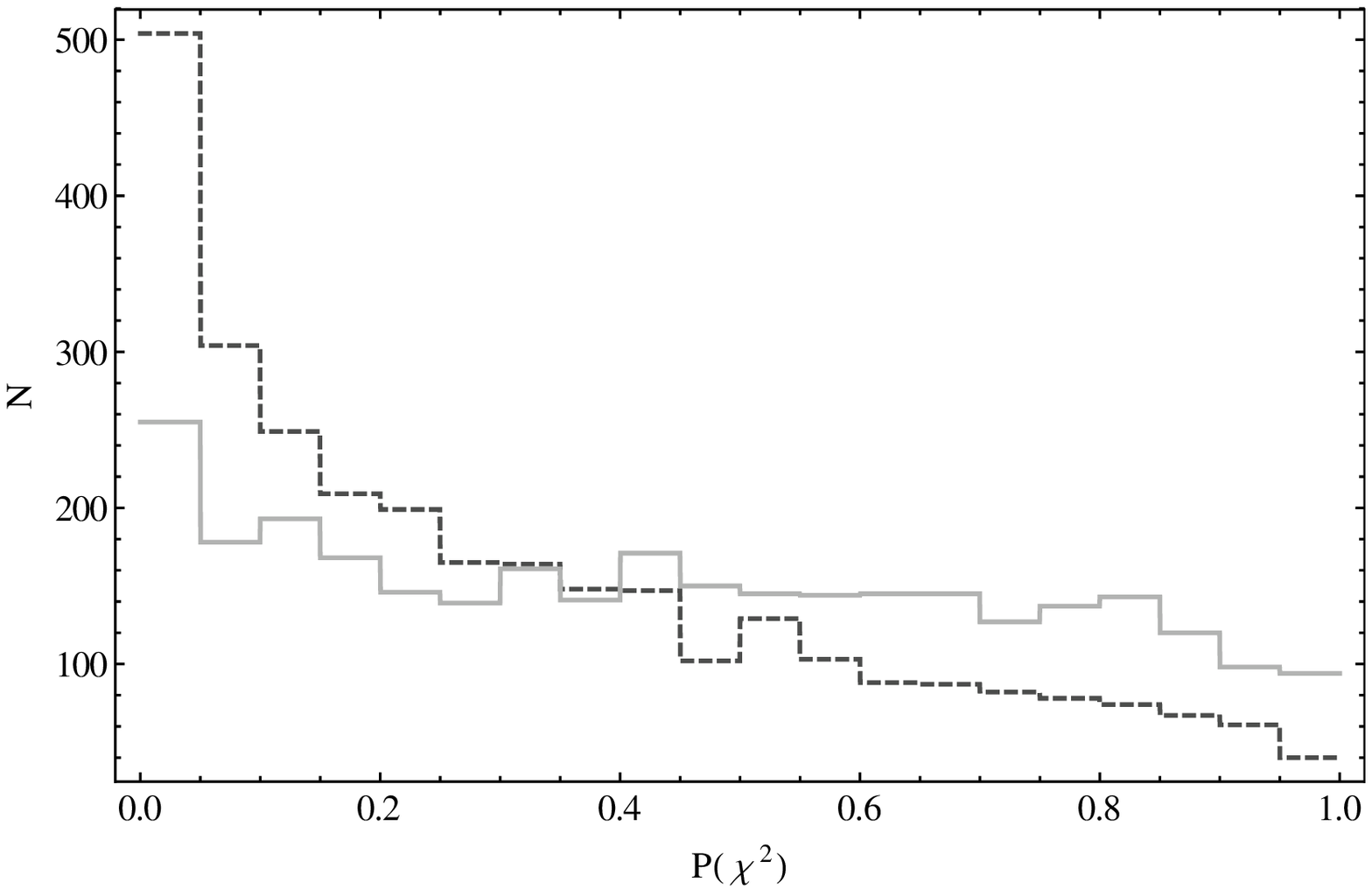}
\caption{Distribution of the $P(\chi^2)$ resulting from the fits to $m_{2\pi}$ in MC samples generated assuming  $1^{++}$ (left) and $2^{-+}$ (right). The solid (dashed) histogram corresponds to the fit to the $2^{-+}$ ($1^{++}$) model.}
\label{fig:toy1}
\end{figure}

With such a tool, we first show the $P(\chi^2)$ values obtained on the $m_{2\pi}$ fits (Fig.~\ref{fig:toy1}) performed by generating the MC events with both the spin hypotheses and by fitting them either with the correct or the wrong hypothesis. These figures show that even when fitting with the incorrect $X$ spin the fit would return a good $P(\chi^2)$. We therefore conclude that $m_{2\pi}$ {\it is not sensitive to the $X$ spin}.

Since we need the invariant mass distributions to perform the combined fit to constraint the $R_J$ variables, we developed 
a more robust method of testing the hypotheses by using as estimator the difference between the $\chi^2$ obtained from the combined fits performed under the two $J$ hypotheses~\cite{Lyons:2008zz}
\begin{equation}
\label{eq:likelihood}
\Delta\chi^2= \chi^2\left(1^{++}\right) - \chi^2\left(2^{-+}\right) 
\end{equation}

\begin{figure}[!h]	
\includegraphics[width=10.0truecm]{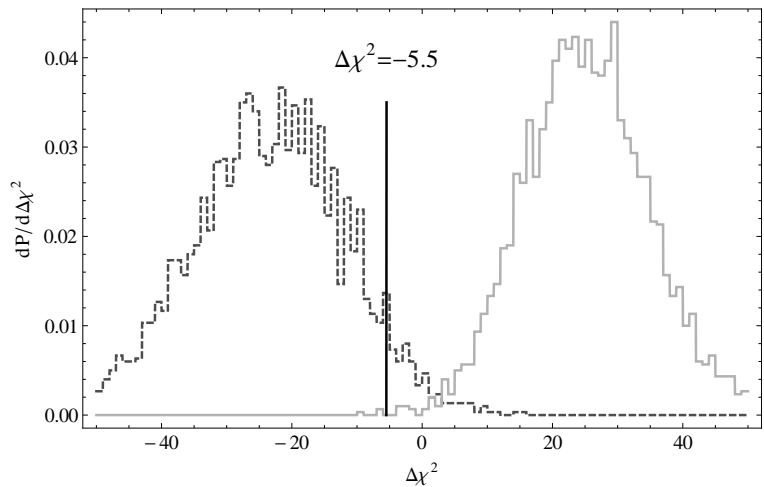}
\caption{Distribution of the $\Delta\chi^2=\chi^2(1^{++})-\chi^2(2^{-+})$  resulting from the combined fits to Monte Carlo data samples  with $n=1$.  The solid (dashed) histogram corresponds to events generated  assuming the $X$ to be a $2^{-+}$ ($1^{++}$) state. We mark with a line
the position of the experimental $\Delta\chi^2$.}
\label{fig:toy2}
\end{figure}

Fig.~\ref{fig:toy2} shows that the distribution of such a variable is peaked around a negative value 
when the MC samples are generated with $1^{++}$ and, viceversa, it is peaked around a positive value when the samples are generated with $2^{-+}$. These distributions are used to calculate the fraction of samples in which $\Delta \chi^2$ has a value larger (for $1^{++}$) or smaller (for $2^{-+}$) than the one obtained on data. This fraction, that we call CL, estimates the probability of the hypothesis.

Given that in the combined fit $\Delta\chi^2=-5.5$, the $2^{-+}$ hypothesis is excluded at 99.9\% C.L., while the $1^{++}$  hypothesis has a C.L. of 5.5\%. The different conclusion with respect to the na\"{i}ve expectations is due to the fact that individual $\chi^2$ values account only for the agreement of the data with the specific hypothesis under test. The $\Delta\chi^2$ approach instead also considers the level of agreement with the other hypothesis, weighting the degrees of freedom with their power to  discriminate among the two hypotheses.
   
\subsection {Fits on sub-samples}
The $\Delta\chi^2$ interpretation of the combined fit shows that even the agreement of the data with the $1^{++}$ hypothesis is marginal and the fit results seem to hint that the disagreement concentrates in the $m_{3\pi}$ distribution (see Fig.~\ref{fig:fitn1}). To quantify this statement we have performed the toy MC analysis on the $\Delta\chi^2$ value obtained separately on the $m_{3\pi}$ distribution (obtained on the $X\to J/\psi\; \pi^+\pi^-\pi^0$ sample) and on the combination of the other distributions (obtained on the $X\to J/\psi\; \pi^+\pi^-$ sample). The results, listed in Tab.~\ref{tab:toyangular}, show that the fits to the $X\to J/\psi\; \pi^+\pi^-$ distributions (second row) exclude the $2^{-+}$ hypothesis and have  an agreement with the  $1^{++}$ hypothesis at 23\% C.L., while conversely the $X\to J/\psi\; \pi^+\pi^-\pi^0$ data (third row) favor the  $2^{-+}$ at 81\% and exclude the $1^{++}$ hypothesis at 99.9\% C.L. The two samples return therefore inconsistent answers.

Finally, in order to compare directly with Ref.~\cite{Hanhart:2011tn} we report in Tab.~\ref{tab:toyangular} also the results performed using exclusively the two invariant mass distributions or the three angular distributions. Also here the former give a clear indication in favor of the 
$2^{-+}$ hypothesis, the latter in favor of the $1^{++}$ hypothesis. The different conclusion with respect to Ref.~\cite{Hanhart:2011tn} comes both from a different theoretical model and from the appropriate treatment of the degrees of freedom with no sensitivity to the spin of the $X$.

\begin{table}
\begin{tabular}{||c|c|c||} \hline
 & $1^{++}$ (dashed curve) & $2^{-+}$ (solid curve) \\ \hline
  \multirow{3}{*}{combined}
& $\chi^2/\text{DOF} = 31.8 / 36$ & $\chi^2/\text{DOF} = 37.3 / 33$\\ 
& $P\left( \chi^2\right) = 67\%$& $P\left( \chi^2\right) = 28\%$\\
& $CL = 5.5\%$& $CL = 0.1\%$ \\ \hline \hline
\multirow{3}{*}{$2\pi$ (angular + mass) } 
& $\chi^2/\text{DOF} = 20.9 / 31$ & $\chi^2/\text{DOF} = 34.7 / 29$\\ 
& $P\left( \chi^2\right)  = 91\%$& $P\left( \chi^2\right) = 21\%$\\
& $CL  = 23\%$& $CL < 0.1\%$ \\ \hline
\multirow{3}{*}{$3\pi$ (mass) }
& $\chi^2/\text{DOF} = 9.9 / 4$ & $\chi^2/\text{DOF} = 1.5 / 3$\\ 
& $P\left( \chi^2\right) = 4\%$& $P\left( \chi^2\right) = 68\%$\\
& $CL  = 0.1\%$& $CL = 81\%$ \\ \hline
\multirow{3}{*}{combined (only mass) }
& $\chi^2/\text{DOF} = 25.2 / 22$ & $\chi^2/\text{DOF} = 17.7 / 20$\\ 
& $P\left( \chi^2\right) = 29\%$& $P\left( \chi^2\right) = 61\%$\\
& $CL  = 0.1\%$& $CL = 46\%$ \\ \hline
\multirow{3}{*}{$2\pi$ (only angular)}
& $\chi^2 / \text{DOF} = 6.6 / 14$ & $\chi^2 / \text{DOF} = 19.6 / 12$\\
& $P\left( \chi^2\right) = 95\%$  & $P\left( \chi^2\right) = 7.6\%$\\
& $CL = 27\%$& $CL < 0.1\%$ \\ \hline
\end{tabular}
\caption{Results of the Toy MC}
\label{tab:toyangular}
\end{table}

\section{Conclusions}

We re-analyzed the $X \to J/\psi \;\pi^+ \pi^- \pi^0$ and $X \to J/\psi \;\pi^+ \pi^- $ invariant mass and angle distributions published by the Belle~\cite{Choi:2011fc} and BaBar~\cite{delAmoSanchez:2010jr} collaborations respectively, with the goal to extract the most information about the spin of the $X$ particle.

With respect to the existing analyzes, we have improved two aspects.
On one side, the $X$ decay amplitudes are parameterized by effective strong couplings, weighting terms  written as products of momenta and polarizations as dictated by Lorentz invariance,  and parity considerations in a model independent way. The strong couplings are given a momentum dependency according to the model defined in Eq.~\eqref{polarfactor}. This requires the introduction of an additional parameter $R$ which can be related to the finite size of hadrons in the interaction region.
The results,\ shown in Tab.~\ref{tab:couplings}, are consistent with $R$ of the order of  1~fm. The treatment is fully relativistic and particularly appropriate for angular analyses.

On the other side, in order to properly account for the sensitivities to the $X$ spin of the considered distributions, we pursue a statistical approach based on Monte Carlo simulations.

We performed $i)$ a global fit based on the whole information available from the $2\pi$, $3\pi$ invariant mass spectra and the angular distributions of the $X\to J/\psi\; \rho$ decays, $ii)$ two separated fits relative to the channels $J/\psi\;\rho$ (where we do the combined fit of the angular and invariant mass distributions) and $J/\psi\;\omega$ (fitting only the invariant mass distribution). 
We also studied the fit to the invariant mass  and the angular  distributions separately.

The combined fit $i)$ excludes the $2^{-+}$ hypothesis at $99.9\%$ C.L., but returns a probability of only  $5.5\%$ of the $1^{++}$ hypothesis being correct. The separate fits $ii)$, return a clear a preference for the $1^{++}$ hypothesis in the $J/\psi \; \rho$ channel with a probability of 23\%  and an 81\% preference for the $2^{-+}$ assignment in the $J/\psi \; \omega$ channel, with very strong exclusions of  other hypotheses.  Such results go in the direction of two different assignments for the two samples, which can lead to a host of interesting considerations.

\section*{Acknowledgements}
We would like to thank F.C.~Porter for the discussion on the statistical approach, and F.~Renga for the analysis of experimental data about the width of the $X$. We also wish to thank C.~Sabelli for her collaboration in the early stages of this work. The work of F. P. was supported by the Research Executive 
Agency (REA) of the European Union under the Grant Agreement 
number PITN-GA-2010-264564 (LHCPhenoNet).
\appendix
\section{\texorpdfstring{\rhoomega}{Rho-omega} mixing}
\label{sec:appendix}
As discussed in Sec.~\ref{sec:widths}, the isosping-violating {\rhoomega} mixing introduces a correction near the pole of the $\omega$, without affecting the core of our analysis.
Moreover, we found that the mixing improves the quality of the fit of $2^{-+}$ and worsens the $1^{++}$. 

\begin{figure}[t]
$\displaystyle{\langle \psi \,\rho |X \rangle =\;}$
\begin{picture}(60,50)(0,23)
\Vertex(35,25){2}
\Line(0,25)(35,25)
\Line(35,25)(60,50)
\Line(35,25)(60,0)
\Text(0,27)[bl]{$\scriptstyle{X}$}
\Text(54,50)[rt]{$\scriptstyle{J/\psi}$}
\Text(52,0)[rb]{$\scriptstyle{\rho}$}
\end{picture} $\displaystyle{\,+\,}$
\begin{picture}(60,50)(0,23)
\Vertex(35,25){2}
\Vertex(48,12){2}
\Line(0,25)(35,25)
\Line(35,25)(60,50)
\Line(35,25)(60,0)
\Text(0,27)[bl]{$\scriptstyle{X}$}
\Text(54,50)[rt]{$\scriptstyle{J/\psi}$}
\Text(53,0)[rb]{$\scriptstyle{\rho}$}
\Text(40,18)[rt]{$\scriptstyle{\omega}$}
\end{picture}
\setlength{\abovecaptionskip}{30pt}
\caption{Feynman diagrams of $X \to J/\psi\,\rho$ including the {\rhoomega} mixing. The second diagram includes the coupling $\epsilon$ and the propagator of the $\omega$. } 
\label{fig:feynman}
\end{figure}
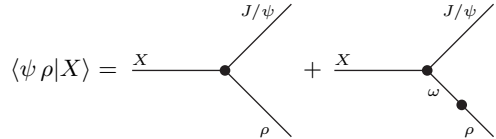

First of all, we have to insert the mixing into \eqref{eq:gamma2pigrezza} and \eqref{eq:gamma3pigrezza}. Following Ref.~\cite{Hanhart:2011tn}, we describe the decay $\omega \to 2\pi$,
as the oscillation $\omega \to \rho \to 2\pi$, as explained in Fig.~\ref{fig:feynman}. We call $-\epsilon$ the coupling of the vertex $\rho\,\omega$, whose value can be extracted by~\cite{Hanhart:2011tn,O'Connell:1995wf}:
\begin{equation}
\epsilon \approx \sqrt{m_\omega\, m_\rho\, \Gamma_\rho \,\Gamma_\omega\, \mathcal{BR}(\omega \to 2\pi)} \approx 3.4 \cdot 10^{-3} \,\textrm{GeV}^2
\end{equation}.

The choice to treat the mixing reproduces naturally the phase of $95^\circ$ of the complex mixing parameter used in Ref.~\cite{Choi:2011fc}.

Under  the hypothesis $1^{++}$ we have
\begin{equation}
 \langle \psi \,\rho(q) |X \rangle = g_{1\psi\rho} \, T - \epsilon \,g_{1\psi\omega} \frac{1}{q^2 - m^2_\omega + i m_\omega \Gamma_\omega} T
\end{equation}
where $T$ is the S-wave scalar described in~\eqref{eq:tensor1}. Similarly for $2^{-+}$
\begin{equation}
 \langle \psi \,\rho(q) |X \rangle  = \left(g_{2\psi \rho} \;T_A + g_{2\psi \rho}^\prime \;T_B\right) - \epsilon\, \frac{g_{2\psi \omega} \;T_A + g_{2\psi \omega}^\prime \;T_B}{q^2 - m_\omega^2 + i m_\omega \Gamma_\omega}
\end{equation}

We can repeat the same argument for $\langle \psi \,\omega(q) |X \rangle$ by swapping the role of $\rho$ and $\omega$. The rest of Eqs.~\eqref{eq:gamma2pi} and \eqref{eq:gamma3pi} (namely the 
Breit-Wigner and the decay into pions) is unchanged, because in this picture it is only the $\rho$ (resp. the $\omega$) which can decay in $2\pi$ ($3\pi$), being the mixing exhausted in the $\langle \psi V |X \rangle$ matrix element.

Moreover, if we take the mixing into account, the parameters $r_{J\rho}$ and $r_{J\omega}$ of the fit appear together in both channels; therefore we must impose their correct relative normalization. This can be
solved by imposing that $\Gamma\left(X \to J/\psi \,3\pi\right) / \Gamma\left(X \to J/\psi \,2\pi\right) = 0.8 \pm 0.3$~\cite{delAmoSanchez:2010jr,review}.

\begin{figure}[!h]
\advance\leftskip0cm
\includegraphics[width=8.4truecm]{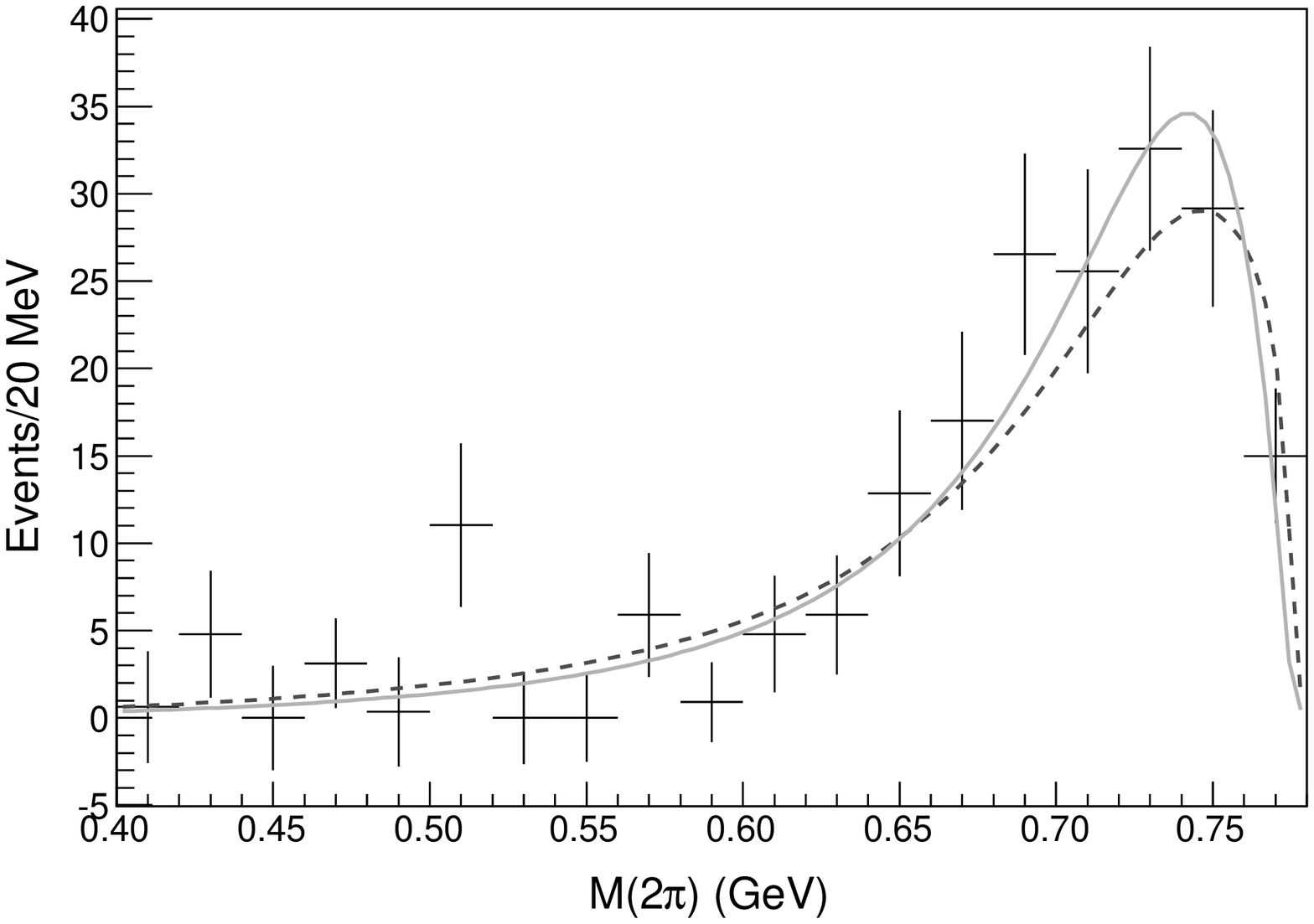}
\includegraphics[width=8.4truecm]{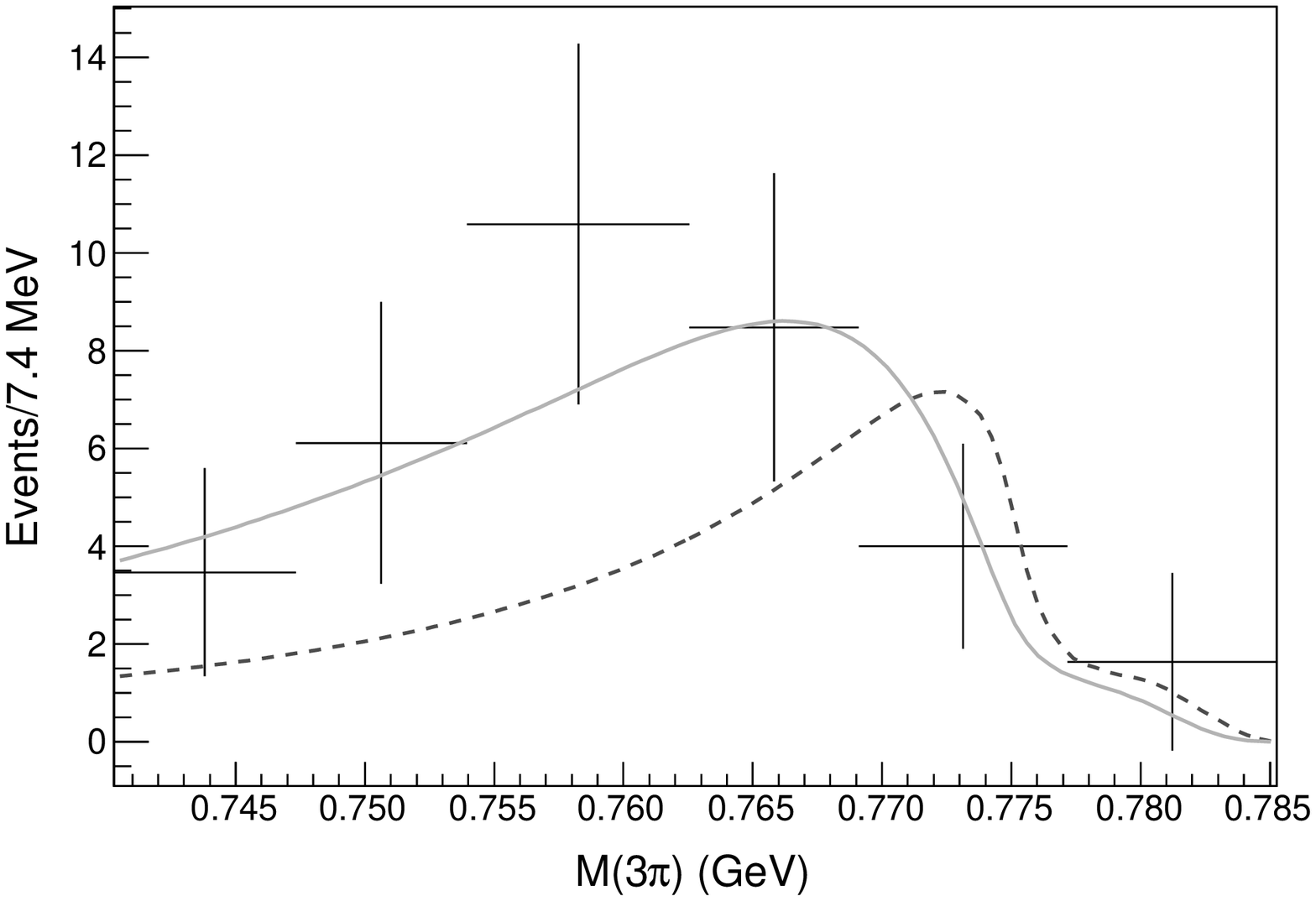}
\caption{Fit including {\rhoomega} mixing to the $m_{2\pi}$ (left) and $m_{3\pi}$ (right) distributions, with the model with $n=1$. The dashed curve refers to the $1^{++}$ hypothesis, the solid one to the $2^{-+}$ one.
}
\label{fig:fitmixing}
\end{figure}

\begin{figure}[!h]	
\includegraphics[width=8.4truecm]{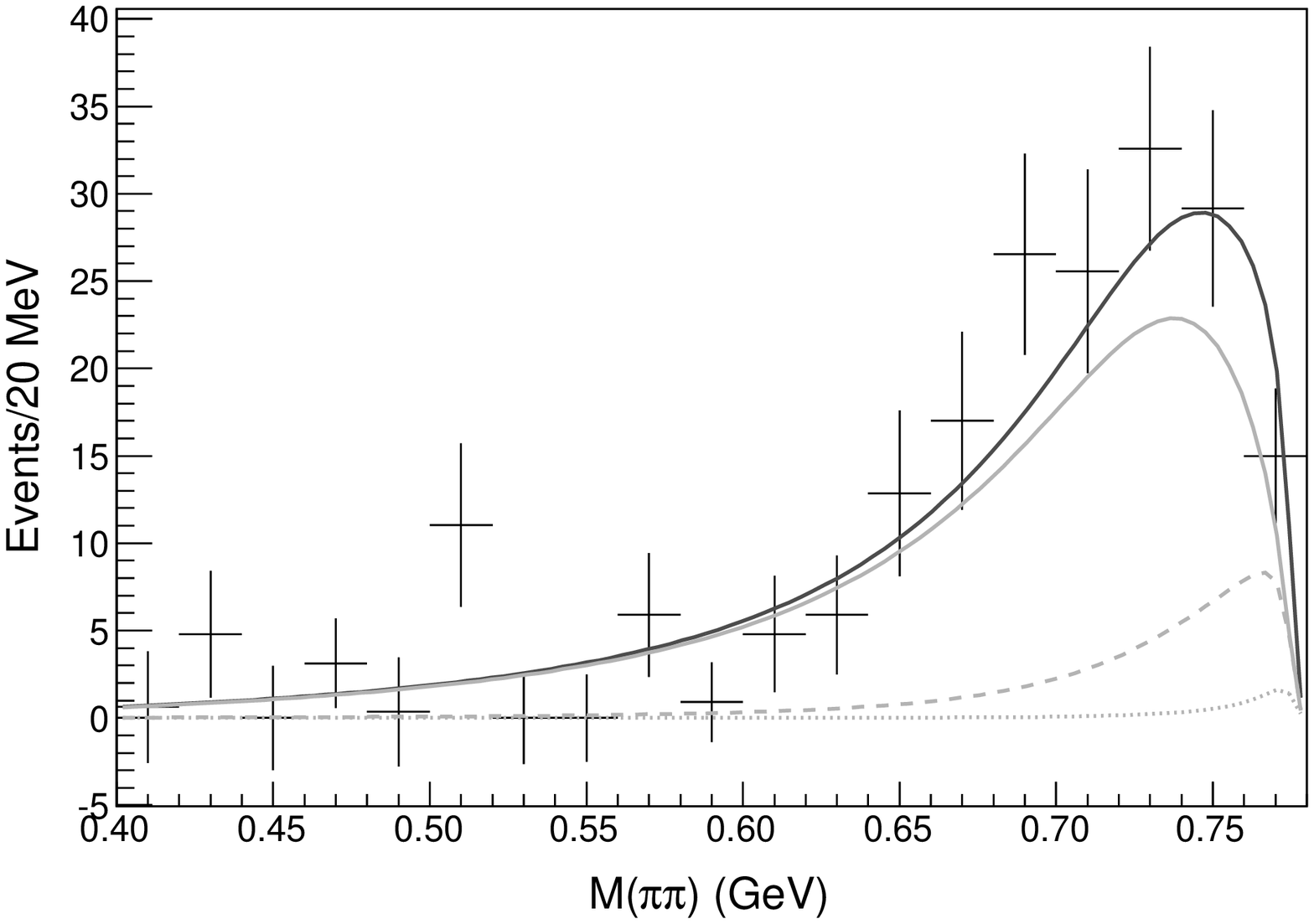}
\includegraphics[width=8.4truecm]{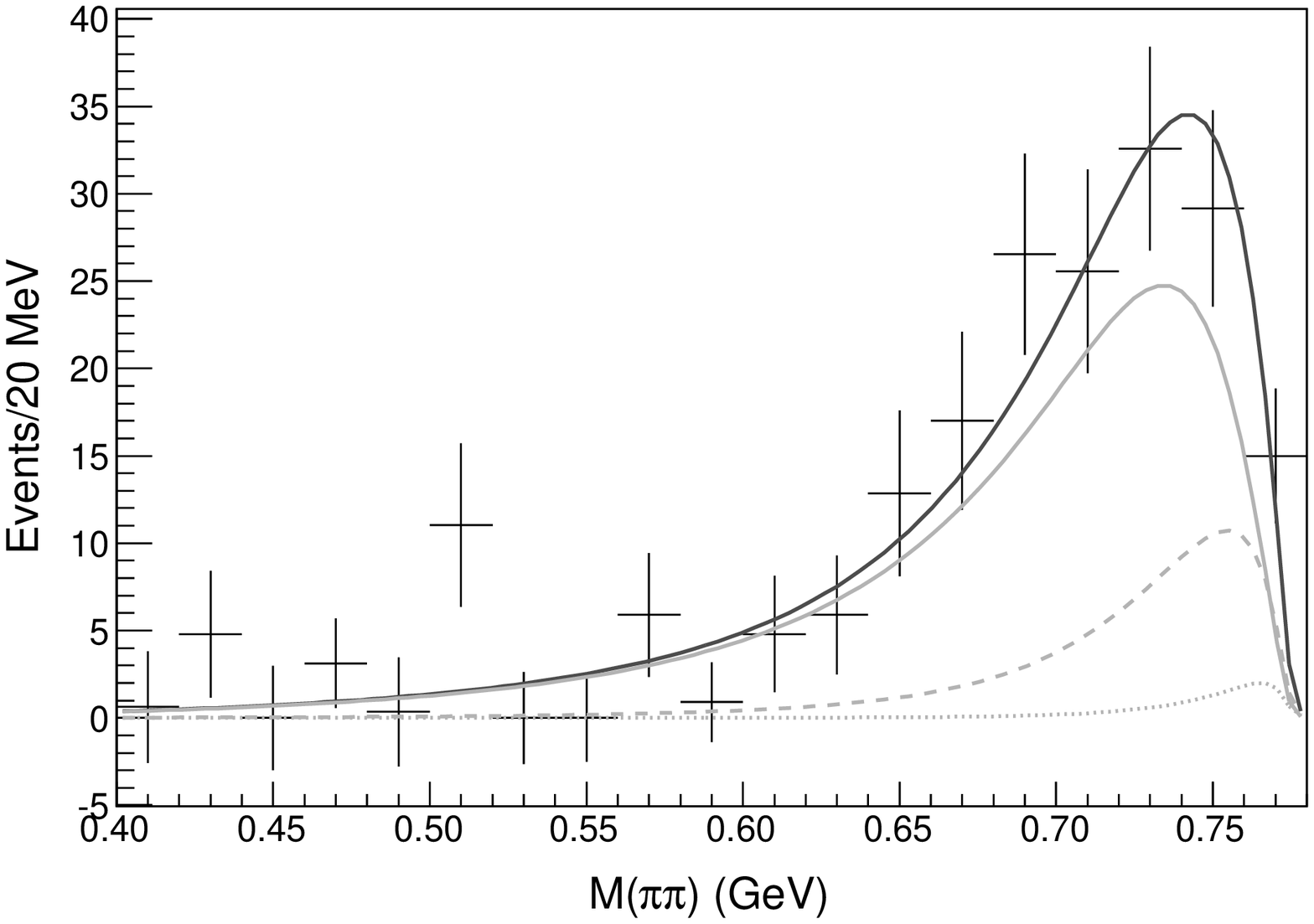}
\caption{Fit to the $m_{2\pi}$ distribution including {\rhoomega} mixing, with the model with $n=1$, and under the hypothesis $J^{PC}=1^{++}$~(left) and $2^{-+}$~(right).
The solid light curve is the $\rho$ contribution to the fit; the dotted light curve is the $\omega$ contribution, and the dashed curve is the interference term. The darker curve is the sum of all contributions.}
\label{fig:splitmixing}
\end{figure}

The angular distributions are unaffected by this term since they depend exclusively on the spin of the particles involved and not
on the invariant mass spectrum. Therefore, to show the impact of the  {\rhoomega} mixing we perform fits to the $m_{2\pi}$  and $m_{3\pi}$  distributions including this effect. The results are shown in Fig.~\ref{fig:fitmixing}, with the individual components detailed for the $m_{2\pi}$ distribution in Fig.~\ref{fig:splitmixing}.  The resulting values of the radii vary from $R_1=1.6\pm 0.3$ to $R_1=1.1\pm 0.4$ and from $R_2=5.6\pm 0.8$ to $R_2=4.5\pm 0.7$. As far as the hypothesis testing is concerned the $\chi^2/\textrm{DOF}$ changes from $25.2/22$ to $29.3/22$ for the $1^{++}$ hypothesis, and from $17.7/20$ to $16.0/20$ for the  $2^{-+}$ one. None of the relevant conclusions is altered by this small effect and we will therefore neglect it in the main analysis.

One could have thought that the presence of the narrow propagator of the $\omega$ would have constrained the fit to raise a peak at $m_\omega = 782$ MeV. The smallness of $\epsilon$, instead, does not allow
this, and the curve stays smooth at the pole, coherently with Refs.~\cite{Abulencia:2005zc,Choi:2011fc,Hanhart:2011tn}.

\section{$n=2$ case}
\label{sec:nequal2}
The theoretical model chosen for the fits described in this paper does not specify the value of $n$ in Eq.~\eqref{polarfactor}, and it is therefore a free parameter of the model itself. Before adopting $n=1$ for the rest of our considerations, we have performed a check on the dependence of the results on the choice of $n$.

\begin{figure}[!h]
\advance\leftskip0cm
\includegraphics[width=8.4truecm]{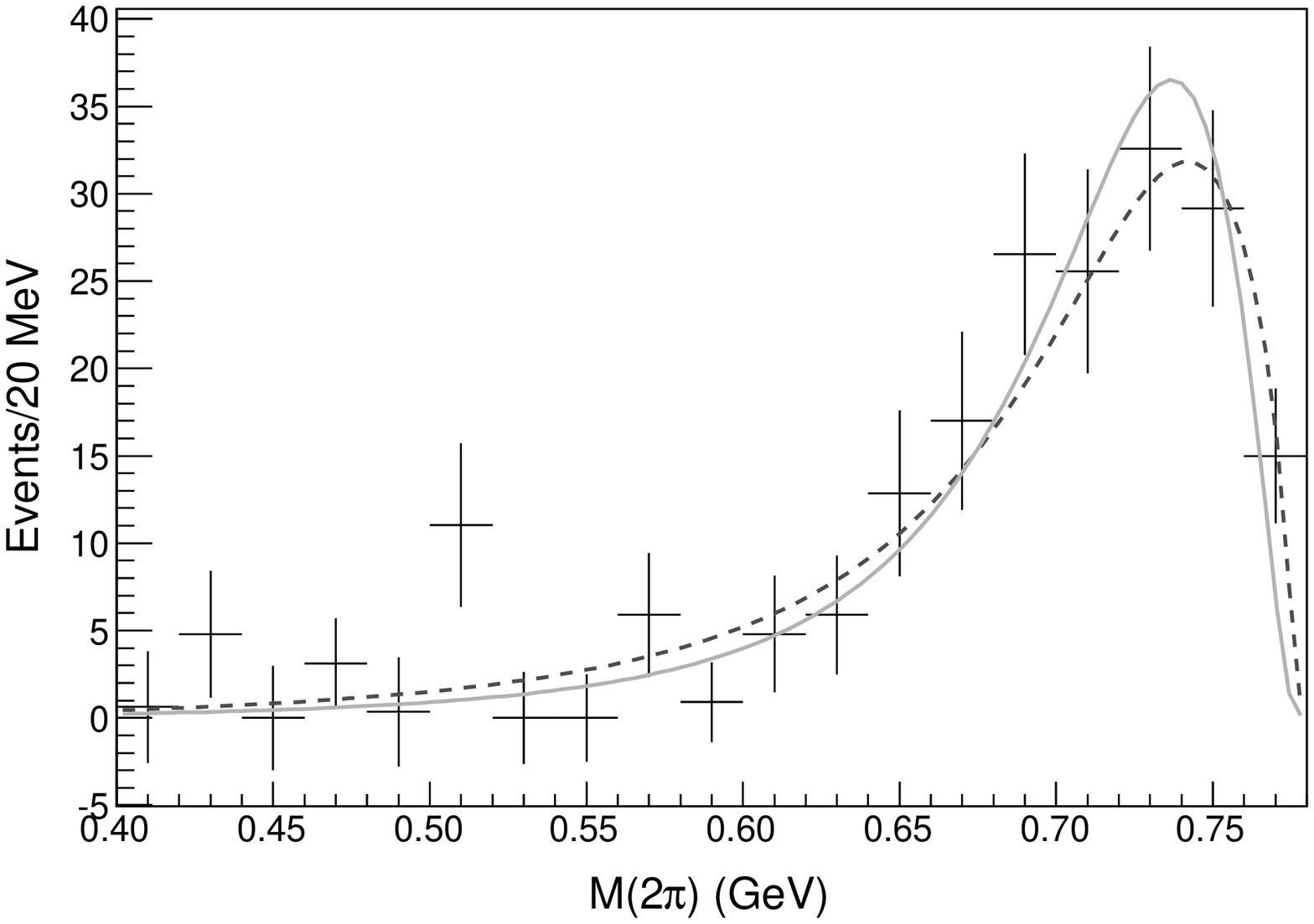}
\includegraphics[width=8.4truecm]{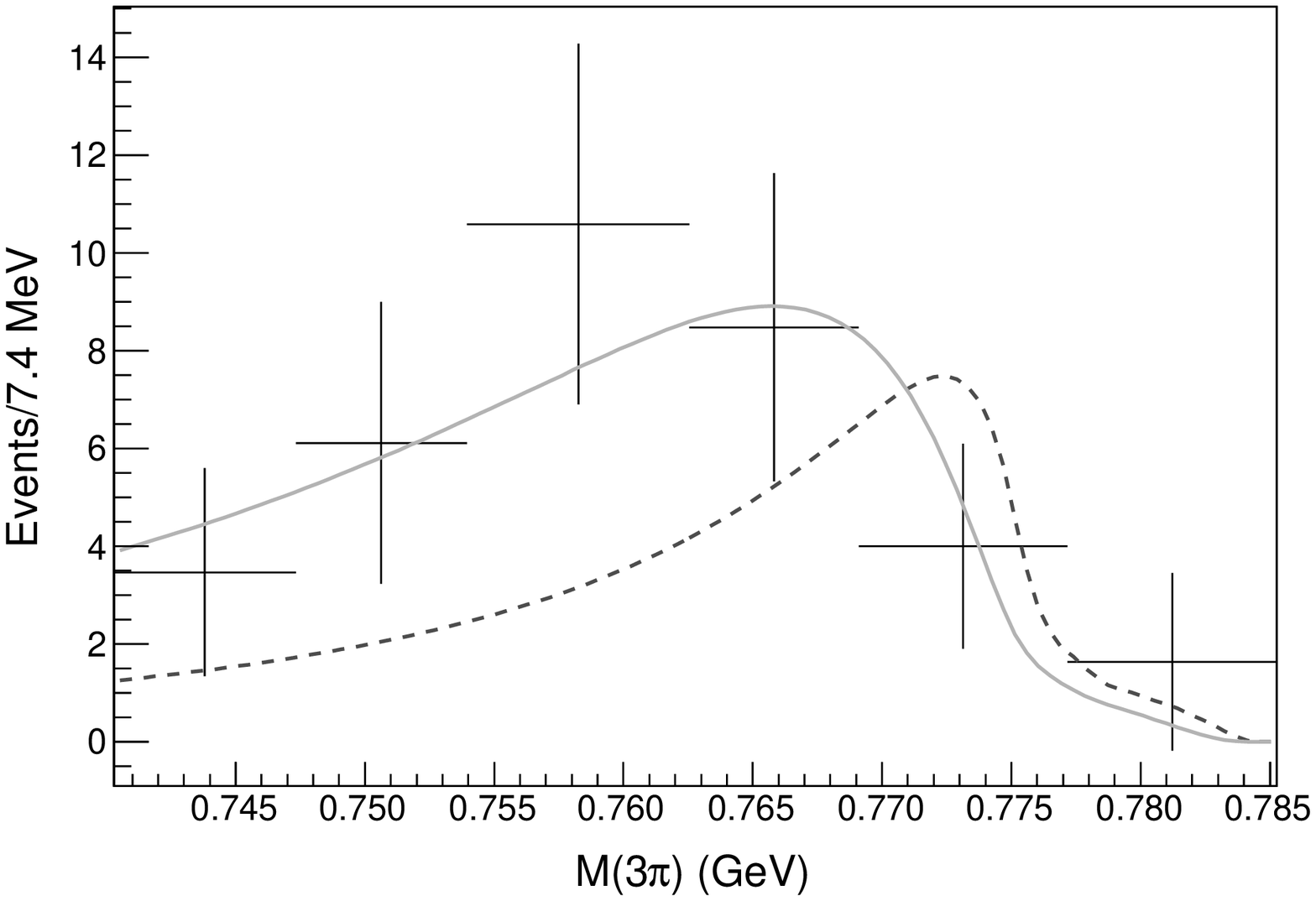}
\caption{Fit to the $m_{2\pi}$ (left) and $m_{3\pi}$ (right) distributions as described in the text, with the model with $n=2$. The dashed curve refers to the $1^{++}$ hypothesis, the solid one is for the $2^{-+}$ one.
}
\label{fig:fitn2}
\end{figure}

Since the angular distributions have shown a scarce dependence on $R$,  we  fitted the invariant mass spectra with $n=2$ comparing the result with the $n=1$ case. The results are shown in Fig.~\ref{fig:fitn2}.
The resulting values of the radii vary from $R_1=1.6\pm 0.3$ to $R_1=1.1\pm 0.2$ and from $R_2=5.6\pm 0.8$ to $R_2=2.8\pm 0.3$. As far as the hypothesis testing is concerned the $\chi^2/\textrm{DOF}$ changes from $25.2/22$ to $24.8/22$ for the $1^{++}$ hypothesis, and from $17.7/20$ to $18.9/20$ for the  $2^{-+}$ one. Also in this case none of the relevant conclusions is altered.

\end{document}